\documentclass{aa}
\usepackage{graphicx,txfonts,natbib}
\bibpunct{(}{)}{;}{a}{}{,}

\begin{document}

\title{Reduced magnetic braking and the magnetic capture model for the formation of ultra-compact binaries}
\author{M.V. van der Sluys \and F. Verbunt \and O.R. Pols }
\titlerunning{Magnetic breaking strength and the magnetic capture model}
\offprints {M.V. van der Sluys, \email{sluys@astro.uu.nl}}

\institute{ Astronomical Institute, Princetonplein 5, NL-3584 CC Utrecht,
            the Netherlands,
            {\tt(sluys@astro.uu.nl)},
            {\tt(verbunt@astro.uu.nl)} and
	    {\tt(pols@astro.uu.nl)}
          }

\date{Received 14 January 2005 / Accepted 13 May 2005}

\abstract{

A binary in which a slightly evolved star starts mass transfer to a neutron star 
can evolve towards ultra-short orbital periods under the influence of magnetic 
braking.  This is called magnetic capture.  In a previous paper we showed that 
ultra-short periods are only reached for an extremely small range of initial binary 
parameters, in particular orbital period and donor mass. Our conclusion was 
based on one specific choice for the law of magnetic braking, and for the loss of mass 
and angular momentum during mass transfer. In this paper we show that for less 
efficient magnetic braking it is impossible to evolve to ultra-short periods, 
independent of the amount of mass and associated angular momentum lost from the binary.

\keywords{Binaries: close, Stars: evolution, Globular clusters: general, X-rays: binaries}
}

\maketitle


\section{Introduction}
\label{sec:intro}

In our previous paper \citep{2005A&A...431..647V} we examined the process of magnetic capture:
a slightly evolved main-sequence star in a binary that transfers mass to a 
neutron-star companion while the orbital period shrinks to the ultra-short-period
regime (less than about 40\,minutes).  
To facilitate comparison with earlier work, we used the same law for
magnetic braking, and the same assumption about the loss of mass and
angular momentum during mass transfer as \citet{2002ApJ...565.1107P}. 
Specifically, we used the law for magnetic braking as
postulated by \citet{1981A&A...100L...7V}, with the extra requirement
that a sufficiently large convective zone is present near the surface
of the star, and we assumed that half of the transferred mass leaves
the binary with the specific angular momentum of the neutron star.  We
concluded that ultra-short periods are reached within the Hubble time
only by binaries within very narrow ranges of initial orbital periods
and donor masses.  In this paper we investigate how this conclusion
changes if we vary the assumptions on the strength of magnetic braking
and on the loss of mass and angular momentum from the system.

Section\,\ref{sec:code} briefly describes the stellar evolution code used
and especially the laws for magnetic braking and system mass loss
that we implemented.  We then show which grids of models were calculated and how the 
statistical study was performed in Sect.\,\ref{sec:distributions}.  The results 
are presented in Sect.\,\ref{sec:results} and discussed in 
Sect.\,\ref{sec:discuss}.  In Sect.\,\ref{sec:concl} we summarise our 
conclusions.


\section{Binary evolution code}
\label{sec:code}

\subsection{The stellar evolution code}

We calculate our models using the \textsc{STARS} binary stellar evolution
code, originally developed by \citet{1971MNRAS.151..351E,1972MNRAS.156..361E} and with updated input
physics as described in \citet{1995MNRAS.274..964P}. Opacity tables are taken from
\textsc{OPAL} \citep{1992ApJ...397..717I}, complemented with low-temperature opacities 
from \citet{1994ApJ...437..879A}.  For more details, see \citet{2005A&A...431..647V}.

\subsection{Angular momentum losses}
\label{sec:amloss}

Loss of angular momentum is essential to shrink the orbit of a binary in which the less massive
star transfers mass to its more massive companion.  We consider three sources of angular momentum loss.

For short periods, gravitational radiation is a strong source of angular momentum loss.
We use the standard description
\begin{equation}
  \frac{dJ_\mathrm{GR}}{dt} ~=~ -\frac{32}{5} \, \frac{G^{7/2}}{c^5} \, \frac{M_1^2 \, M_2^2 \, \left(M_1 + M_2\right)^{1/2}}{a^{7/2}}
  \label{eq:gravrad}
\end{equation}
\citep{pet84}.

The second mechanism of angular momentum loss from the system is by non-conservative mass
transfer.  We assume that only a fraction $\beta$ of the transferred mass is accreted by the
neutron star. The remainder is lost from the system, carrying away a fraction 
$\alpha$ of the specific orbital angular momentum of the neutron star:
\begin{equation}
  \frac{dJ_\mathrm{ML}}{dt} ~=~ - \alpha\left(1-\beta\right) a_1^2 \, \omega \, \dot{M}_2,
  \label{eq:am_ml}
\end{equation}
where $a_1$ is the orbital radius of the neutron star and $\omega$ is the angular velocity.

To keep the models simple, we applied no regular stellar wind to our models, so
that all mass loss from the system and associated angular momentum loss result from
the non-conservative mass transfer described above.

The third source of angular momentum loss in this study is magnetic braking.  
\citet{1981A&A...100L...7V} postulated a law for magnetic braking
\begin{equation}
  \frac{dJ_\mathrm{MB}}{dt} ~=~ -3.8 \times 10^{-30} \, \eta \, M \, R^4 \, \omega^3 \, \mathrm{dyn\,cm} ,
  \label{eq:magnbrak}
\end{equation}
on the basis of the observations by \citet{1972ApJ...171..565S} that the equatorial
rotation velocity $v_e$ of main-sequence G stars decreases with the age
$t$ of the star as $v_e\propto t^{-0.5}$. In our previous paper we
assumed $\eta=1$, after \citet{1983ApJ...275..713R}.
More recent measurements of rotation velocities of stars in the Hyades and Pleiades, however, 
show that M stars have a wide range of rotation
velocities that is preserved as they age \citep{2000AJ....119.1303T}. This indicates
that magnetic braking is less strong for low mass stars than assumed in Eq.\,\ref{eq:magnbrak} with $\eta=1$.
Also, observational evidence indicates that coronal and chromospheric
activity, and with it magnetic braking, saturate to a maximum level at rotation
periods less than about 3\,days (e.g.\ Vilhu 1982\nocite{1982A&A...109...17V}; 
Vilhu \&\ Rucinski 1983\nocite{1983A&A...127....5V}).  
\citet{1984MNRAS.209..227V} showed that to explain a braking with
the strength of Eq.\,\ref{eq:magnbrak} for a solar-type star, the star must have a magnetic 
field in excess of $\sim200$\,G for a slow rotator, and in addition a stellar wind loss in 
excess of $5\times10^{-10}M_\odot$/yr for a fast rotator (for which the corotation velocity 
of the wind matter is much higher than the escape velocity -- see also \citet{1981ApJ...245..650M}). 
A smaller field or less wind (for the fast rotator) automatically leads to a lower braking.

Many theoretical descriptions of angular momentum 
loss due to a magnetized wind can be found in the literature (among others Kawaler 1988\nocite{1988ApJ...333..236K}; 
Stepien 1995\nocite{1995MNRAS.274.1019S};  Eggleton \& Kiseleva-Eggleton 2002\nocite{2002ApJ...575..461E}; 
Ivanova \& Taam 2003\nocite{2003ApJ...599..516I}).  These prescriptions depend 
on properties of the star (for instance wind mass loss rate, magnetic field strength, 
corona temperature) that are poorly known from observations for main-sequence stars and even less for evolving stars.  These angular momentum 
prescriptions vary in strength and dependence on the stellar parameters.  We 
have selected two different semi-empirical prescriptions to investigate the effect of 
reduced braking on the mechanism of magnetic capture.  In Sect.\,\ref{sec:discuss} we 
will show that these two different implementations of magnetic braking dominate the evolution of the binary 
in two completely different phases of their life.

First, we retain the functional dependence of the braking on stellar
mass and radius given by Eq.\,\ref{eq:magnbrak}, but arbitrarily reduce the strength
by taking $\eta=0.25$ (reduced braking) and $\eta=0$ (no braking).
Second, we use a new law for magnetic braking, derived on the basis of
the ranges of rotation velocities in the Hyades and Pleiades, which
includes saturation at a critical angular rotation velocity
$\omega_\mathrm{crit}$ \citep{2000ApJ...534..335S}:

\[
  \frac{dJ_\mathrm{MB}}{dt} ~=~ - K \, \left(\frac{R}{R_\odot}\right)^{0.5} \left(\frac{M}{M_\odot}\right)^{-0.5}\, \omega^3, ~~~ \omega \leq \omega_\mathrm{crit} 
\]
\begin{equation}
  \frac{dJ_\mathrm{MB}}{dt} ~=~ - K \, \left(\frac{R}{R_\odot}\right)^{0.5} \left(\frac{M}{M_\odot}\right)^{-0.5}\, \omega \, \omega_\mathrm{crit}^2 ,  ~ \omega > \omega_\mathrm{crit}
\label{eq:satMB}
\end{equation}

From \citet{2003ApJ...582..358A} we take the value $K = 2.7 \times 10^{47}$\,g\,cm$^2$\,s that 
reproduces the angular velocity of the Sun at the age of the Sun.
\citet{1997ApJ...480..303K} require a mass-dependent value for $\omega_\mathrm{crit}$  
and they scale this quantity inversely with the turnover timescale for the convective 
envelope $\tau_\mathrm{to}$ of the star at an age of 200\,Myr:
\begin{equation}
  \omega_\mathrm{crit} ~=~ \omega_\mathrm{crit,\odot}  \, \frac{\tau_\mathrm{to,\odot}}{\tau_\mathrm{to}}
\label{eq:ocrit}
\end{equation}
They use a fixed value for $\omega_\mathrm{crit}$, because they consider main-sequence stars and 
the value of $\tau_\mathrm{to}$ does not change much during this evolution period.  However, we 
consider donor stars in a binary system that change substantially during their evolution
and hence use the instantaneous value for $\tau_\mathrm{to}$.  This convective turnover timescale 
is determined by the evolution code by integrating the inverse velocity of convective cells, as given by the
mixing-length theory \citep{1958ZA.....46..108B}, over the
radial extent of the convective envelope. We further use a value of $\omega_\mathrm{crit,\odot} = 2.9 \times 10^{-5}$\,Hz, 
equivalent to $P_\mathrm{crit,\odot} = 2.5$\,d 
(Sills et al. 2000\nocite{2000ApJ...534..335S} find that a value for $\omega_\mathrm{crit,\odot}$ of around 10 
times the current solar angular velocity is needed to reproduce observational data of young clusters
with a rigidly rotating model), and 
$\tau_\mathrm{to,\odot} = 13.8$\,d, the value that the evolution code gives for a $1.0\,M_\odot$ 
star at the age of 4.6\,Gyr.  

In both prescriptions (Eqs.\,\ref{eq:magnbrak} and \ref{eq:satMB}) we follow 
\citet{2002ApJ...565.1107P} and reduce the magnetic braking by an ad hoc term 
\begin{equation}
  \exp(1 - 0.02/q_\mathrm{conv}) ~~~~  \mathrm{if} ~ q_\mathrm{conv} \, < \, 0.02,
\label{eq:qconv}
\end{equation}
where $q_\mathrm{conv}$ is the fractional mass of the convective envelope.
In this way we account for the fact that stars with a small or no convective mantle do not
have a strong magnetic field and will therefore experience little or no magnetic braking.
Notice that Eq.\,\ref{eq:ocrit} alone predicts that stars with higher mass have a higher $\omega_\mathrm{crit}$,
because they have a higher surface temperature, hence a smaller convective mantle and a shorter $\tau_\mathrm{to}$.
The application of the term in Eq.\,\ref{eq:qconv} prevents that these stars experience unrealistically strong magnetic braking.


\section{Creating theoretical period distributions}
\label{sec:distributions}

\subsection{Binary models}
\label{sec:models}

Using the binary evolution code described in Sect.\,\ref{sec:code}, with the non-saturated
magnetic-braking law of Eq.\,\ref{eq:magnbrak} we calculated
grids of models for Z=0.01, the metallicity of the globular cluster NGC\,6624, and Y=0.26.  We choose 
initial masses between 0.7 and 1.5\,$M_\odot$ with steps of 0.1\,$M_\odot$.  For each mass  
we calculated models with initial periods (P$_\mathrm{i}$) between 0.5 and 2.5 days with steps of 0.5\,d
for all masses and dropped the lower limit for the initial period where necessary, down 
to 0.25\,d.  Around the bifurcation period between converging and diverging systems, 
where the shortest orbital periods occur, we narrow the steps in P$_\mathrm{i}$ to 0.05 or 
sometimes even 0.02\,d.  

Another series of models was calculated with a similar grid of initial masses and periods,
but with the magnetic-braking law of Eq.\,\ref{eq:satMB} that includes saturation of the magnetic field
strength at high angular velocities.

\subsection{Statistics}
\label{sec:statistics}

In order to create a theoretical period distribution for a population of stars, we proceed as described in
\citet{2005A&A...431..647V}.  First, we draw a random initial period (P$_\mathrm{i}$) and calculate the 
time-period track for this P$_\mathrm{i}$ by interpolation from the two bracketing calculated 
tracks.  Second, we pick a random moment in time and interpolate within the obtained
time-period track to get the orbital period of the system at that moment in time.
The system is accepted if mass transfer is occurring and the period derivative is negative.
The details of this interpolation are described in \citet{2005A&A...431..647V}.  We do this $10^6$
times for each mass to produce a theoretical orbital-period distribution for a given initial mass and
given ranges in log\,P$_\mathrm{i}$ and time.

To simulate the period distribution for a population of stars with an initial mass
distribution, we add the distributions of different masses.  
In \citet{2005A&A...431..647V}, we show that the result depends very little on the 
weighting, so that we simulate a flat distribution in initial mass.


\section{Results}
\label{sec:results}

\subsection{Reduced magnetic braking}
\label{sec:mbstrength}

\begin{figure}
\resizebox{\hsize}{!}{\rotatebox{-90}{\includegraphics{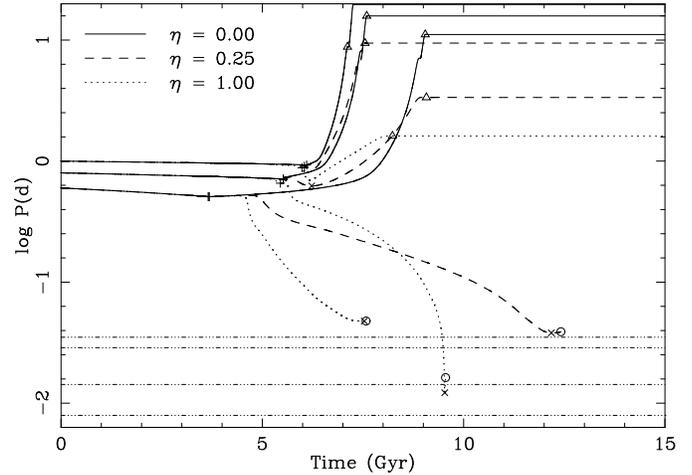}}}
 \caption{
  Time-period tracks for $Z = 0.01$, $M_\mathrm{i} = 1.1$\,M$_\odot$ with $P_\mathrm{i}$ = 
  0.6\,d, $P_\mathrm{i}$ = 0.8\,d, and 
  $P_\mathrm{i}$ = 1.0\,d.  Each model is shown for three
  different values of  $\eta$: 0.0 (solid lines), 0.25 (dashed lines) and 1.0 (dotted lines).
  The symbols mark special points in the evolution: $+$ marks the start of Roche-lobe 
  overflow (RLOF), $\times$ the minimum period, $\bigtriangleup$ the end of RLOF
  and O marks the end of the calculation.  The four dash-dotted horizontal lines show the
  orbital periods of the closest observed LMXBs in globular clusters: 11.4 and 20.6, and
  in the galactic disk: 41 and 50 minutes.
  \label{fig:tracks_allmb} }
\end{figure}

We have calculated three grids of models as described in Sect.\,\ref{sec:models} with the 
non-saturated magnetic-braking law given by Eq.\,\ref{eq:magnbrak}. We have given the three 
grids different braking strengths by changing the value for $\eta$.  We used the values $\eta = 1.0$ 
(as in van der Sluys et al. 2004\nocite{2005A&A...431..647V}), $\eta = 0.25$ and $\eta = 0.0$.  For the last set, there is no 
magnetic braking and the angular momentum loss comes predominantly from gravitational 
radiation.  For all models in these grids, half of the transferred mass is ejected from the
system with the specific angular momentum of the neutron star, i.e. we used $\alpha = 1$ 
and $\beta = 0.5$ in Eq.\,\ref{eq:am_ml}. Figure\,\ref{fig:tracks_allmb} shows time-period tracks 
for models from the three grids with selected initial orbital periods and 
$M_\mathrm{i} = 1.1\,M_\odot$.

The figure shows clearly that initially similar models evolve in different ways, but only
after mass transfer has started.  
This is because a low-metallicity main-sequence 1.1\,$M_\odot$ star has a high surface 
temperature, hence a small convective envelope ($q_\mathrm{conv} \approx 10^{-3}$) and therefore 
effectively no magnetic braking at that stage (see Eq.\,\ref{eq:qconv}).
After mass transfer starts, the surface
temperatures drop and the differences in magnetic braking strength become apparent.
It is obvious that a model that experiences weaker magnetic braking may diverge where a similar
model with stronger braking converges, and that models with weak magnetic braking converge slower 
than models with strong magnetic braking.

\begin{figure}
\resizebox{\hsize}{!}{\rotatebox{-90}{\includegraphics{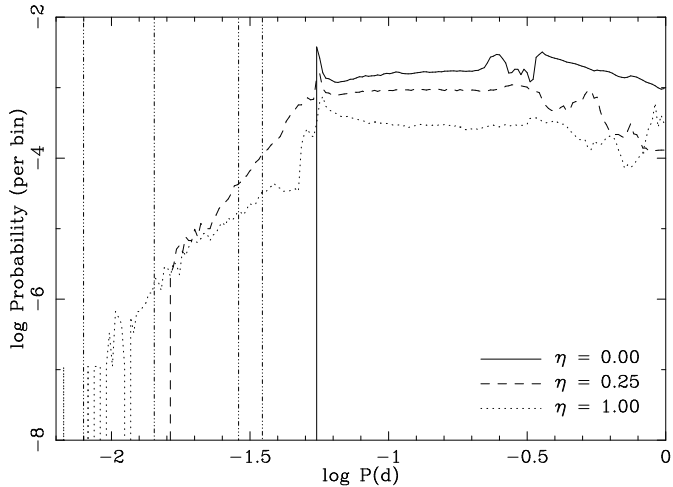}}}
 \caption{
  Period distributions for the magnetic capture model for 
  $\eta$ = 0.0 (solid lines), 0.25 (dashed lines) and 1.0 (dotted lines).
  It is clear that the cut-off for lower orbital periods strongly depends on the
  strength of the magnetic braking.  The vertical axis displays the logarithm of 
  the probability that an X-ray binary with a certain orbital period is found.
  The four vertical dash-dotted lines show the same observed orbital periods as the horizontal
  lines in Fig.\,\ref{fig:tracks_allmb}.  The probability is computed by distributing the
  accepted periods into bins of width $\Delta \log P = 0.011$ and dividing the number in each bin
  by the total number of systems. 
  \label{fig:phist_allmb}  }
\end{figure}

For each grid of models we produce a statistical sample as explained in Sect.\,\ref{sec:statistics}.  
The results are period distributions for three populations of stars with initial masses
between 0.7 and 1.5\,$M_\odot$ and ages between 10 and 13\,Gyr.  The distributions
are compared in Fig.\,\ref{fig:phist_allmb}.

The most striking difference in the period distributions for the three values of $\eta$ is 
the shortest orbital period produced in the magnetic capture model.  
In models with reduced magnetic braking the orbits do not converge to ultra-short periods 
before the Hubble time, and the cut-off at the low-period end of the distribution accordingly
lies at a higher period. This is also
the reason why there are more systems with orbital periods of around 0.1\,d for $\eta = 0.0$ 
than for $\eta = 1.0$; the missing models with stronger braking have already converged to lower 
orbital periods, or beyond the period minimum.

\subsection{Saturated magnetic braking}
\label{sec:satmb}

We have calculated one grid of models described in Sect.\,\ref{sec:models} with the 
saturated magnetic-braking law given by Eq.\,\ref{eq:satMB}. 
In this prescription the magnetic field saturates at a certain
critical angular velocity $\omega_\mathrm{crit}$, that depends on the convective turnover timescale 
of the donor star, as shown in Eq.\,\ref{eq:ocrit}.  At an angular velocity higher than 
$\omega_\mathrm{crit}$, the magnetic braking scales linearly with $\omega$ rather than cubically.  
As the typical  initial critical spin {\em period} is a few days, this is long compared to the 
initial orbital and  -- since the spins and orbits of our models are generally synchronised -- 
spin period, and therefore replacing the prescription of Eq.\,\ref{eq:magnbrak} by that of 
Eq.\,\ref{eq:satMB} can be expected to have an effect similar to lowering the strength of the
magnetic braking, as we did in Sect.\,\ref{sec:mbstrength}.  Because we will see in Sect.\,\ref{sec:massloss} that the
shortest orbital periods are reached for models with conservative mass transfer, all models in 
this grid have $\beta = 1.0$ in Eq.\,\ref{eq:am_ml}.

\begin{figure}
\resizebox{\hsize}{!}{\rotatebox{-90}{\includegraphics{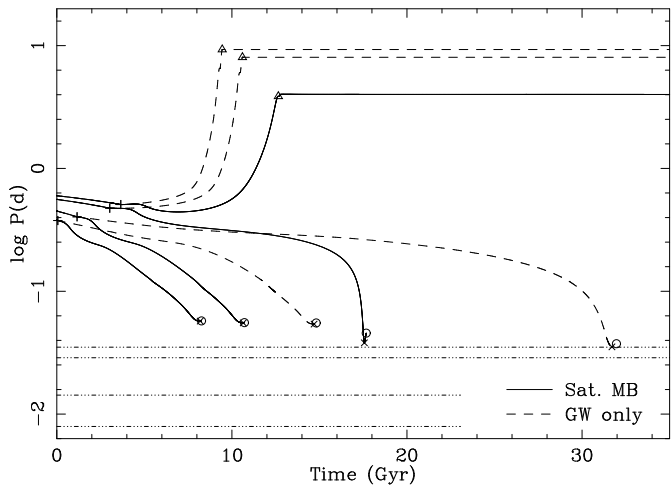}}}
 \caption{
  Time-period tracks for $Z = 0.01$, $M_\mathrm{i} = 1.1$\,M$_\odot$ with $P_\mathrm{i}$ = 
  0.38\,d (the shortest possible $P_\mathrm{i}$ for this model), $P_\mathrm{i}$ = 0.45\,d, 
  $P_\mathrm{i}$ = 0.56\,d, and $P_\mathrm{i}$ = 0.60\,d.  
  Each model is shown for a magnetic braking law according to
  Eq.\,\ref{eq:satMB} (Sat. MB, solid lines) and no magnetic braking, but gravitational waves only
  (GW only, dashed lines).  The symbols and horizontal dash-dotted lines are as in Fig.\,\ref{fig:tracks_allmb}. 
  Note that the time axis extends far beyond the Hubble time.
  \label{fig:tracks_sills_gw} }
\end{figure}

Figure\,\ref{fig:tracks_sills_gw} compares the tracks of $1.1 M_\odot$ models from this grid with 
tracks taken from Sect.\,\ref{sec:massloss} with conservative mass transfer and without magnetic braking, 
i.e. $\beta = 1.0$ and $\eta = 0.0$.
We see similar differences between the two sets of models as seen in Fig.\,\ref{fig:tracks_allmb}, but
the magnetic braking is clearly too weak to evolve the systems to less than 75\,min within the
Hubble time.

\begin{figure}
\resizebox{\hsize}{!}{\rotatebox{-90}{\includegraphics{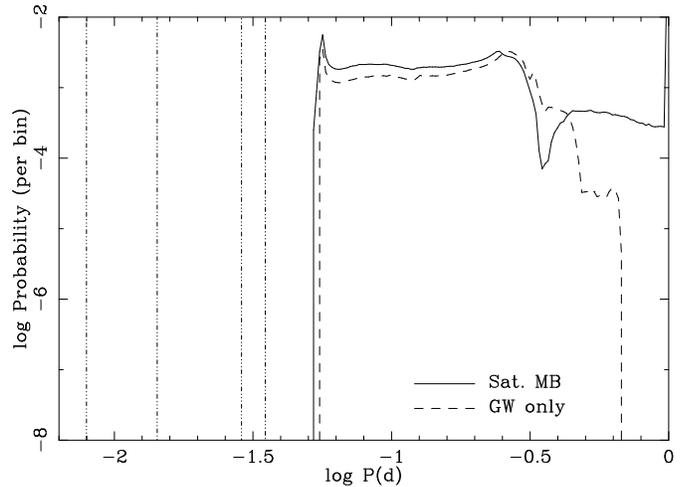}}}
 \caption{
  Period distribution for the magnetic capture model using the magnetic braking law described in Eq.\,\ref{eq:satMB}
  (Sat. MB, solid line) compared to the period distribution for models without braking, but with gravitational
  waves only (GW only, dashed line).
  The four vertical dash-dotted lines show the same observed orbital periods as the horizontal
  lines in Fig.\,\ref{fig:tracks_allmb}.  The probability is calculated in the same way as in Fig.\,\ref{fig:phist_allmb}.
  \label{fig:phist_sills_gw}  }
\end{figure}

We performed statistics on the model as described in Sect.\,\ref{sec:statistics}; the result is
displayed in Fig.\,\ref{fig:phist_sills_gw} and compared to the period distribution for a grid
of models with $\beta = 1.0$ and $\eta = 0.0$.  

\subsection{The influence of mass loss}
\label{sec:massloss}

\begin{figure}
\resizebox{\hsize}{!}{\rotatebox{-90}{\includegraphics{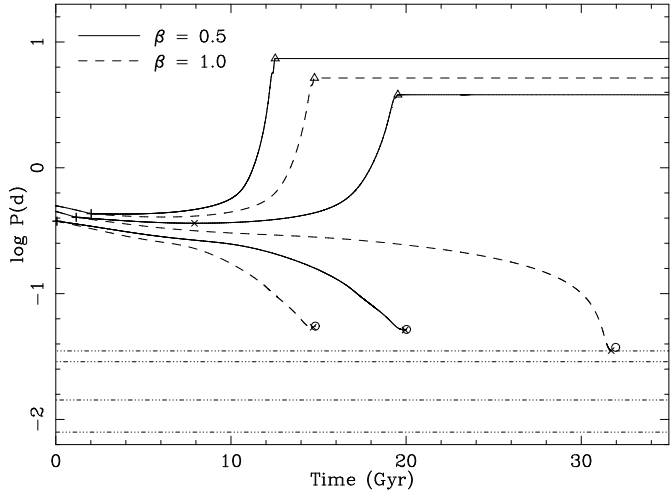}}}
 \caption{
  Time-period tracks for $Z = 0.01$, $M_\mathrm{i} = 1.1$\,M$_\odot$ with $P_\mathrm{i}$ = 
  0.38\,d (the shortest possible $P_i$), $P_\mathrm{i}$ = 0.45\,d and 
  $P_\mathrm{i}$ = 0.50\,d.  Each model is shown for two
  different values of $\beta$ ($\beta = 0.5$, solid lines and $\beta = 1.0$, dashed lines) 
  and has no magnetic braking ($\eta = 0.0$).  
  The symbols and horizontal dash-dotted lines are as in Fig.\,\ref{fig:tracks_allmb}.
  \label{fig:tracks_accr_gw}  }
\end{figure}

\begin{figure}
\resizebox{\hsize}{!}{\rotatebox{-90}{\includegraphics{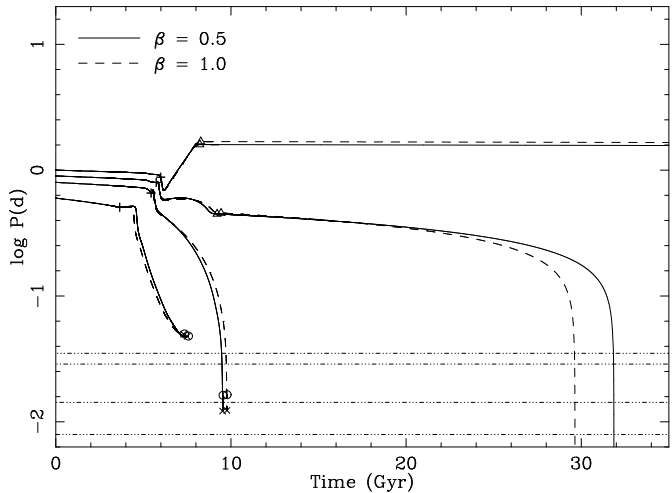}}}
 \caption{
  Time-period tracks for $Z = 0.01$, $M_\mathrm{i} = 1.1$\,M$_\odot$ with $P_\mathrm{i}$ = 
  0.6\,d, $P_\mathrm{i}$ = 0.8\,d, $P_\mathrm{i}$ = 0.9\,d, and 
  $P_\mathrm{i}$ = 1.0\,d.  Each model is shown for two
  different values of $\beta$ ($\beta = 0.5$, solid lines and $\beta = 1.0$, dashed lines) 
  and has full magnetic braking ($\eta = 1.0$).  
  The symbols and horizontal dash-dotted lines are as in Fig.\,\ref{fig:tracks_allmb}.
  \label{fig:tracks_accr_mb}  }
\end{figure}

In our previous paper we have assumed that half of the transferred mass in our models is lost by the accretor
and leaves the system with the specific angular momentum of the accretor.  To see what influence
this assumption has on the results of our study, we calculated a number of models with conservative
mass transfer, so that $\beta = 1.0$ in Eq.\,\ref{eq:am_ml}.  We calculated two sets of conservative
models, one set without magnetic braking ($\eta = 0$ in Eq.\,\ref{eq:magnbrak}) and one set with
full braking ($\eta = 1$).  The time-period tracks of selected models are compared to previous models
with $\beta = 0.5$ in Figs.\,\ref{fig:tracks_accr_gw} and \ref{fig:tracks_accr_mb}.

Figure\,\ref{fig:tracks_accr_gw} shows that the time-period tracks of models with gravitational waves
as the dominant angular momentum loss source are changed noticeably by a change in $\beta$. Converging models
reach their minimum period much earlier for conservative models than for non-conservative models.  
The reason for this is that mass loss from the binary according to Eq.\,\ref{eq:am_ml} leads to a widening
of the binary for the value of $\alpha$ we use.  However,
even for the shortest possible initial period (0.38\,d), and therefore the earliest possible
period minimum for these systems, the time of the minimum shifts from 19.9\,Gyr to 14.7\,Gyr with a 
period of 78\,min.  The conclusion is that this effect makes no difference to the number or distribution
of ultra-compact binaries.

For models with magnetic braking, the differences between the two sets of models is much smaller, as shown in 
Fig.\,\ref{fig:tracks_accr_mb}.  The reason for this is that the orbital evolution is completely
dominated by the strong magnetic braking, so that changes in less important terms, like the amount of 
mass loss from the system and the associated angular momentum loss, are of very little influence.
The models with full magnetic braking {\em can} produce ultra-compact binaries within the Hubble time;
the distribution of ultra-short periods in these models is slightly affected by a change in $\beta$ 
(see Fig.\,\ref{fig:tracks_accr_mb}), but not enough to change the overall conclusion of \citet{2005A&A...431..647V}.

\section{Discussion}
\label{sec:discuss}

It is clear that the magnetic capture scenario to create ultra-compact binaries depends very strongly
on the strength of the magnetic braking used.  By simply scaling down the \citet{1981A&A...100L...7V} prescription for magnetic braking, the 
results are, as can be expected intuitively,

\begin{itemize}
\item The bifurcation period between converging and diverging systems decreases, which means that only
models with a lower initial orbital period will converge.

\item The rate at which a system converges is lower, so that minimum periods are reached at a later
time.  This can imply that ultra-compact periods occur only after a Hubble time.

\item Because reaching the minimum period takes much longer, a small offset in the initial period has
much more impact on the evolution of the system.  Because of this, the initial period range that leads 
to ultra-compact systems for a certain initial mass is much smaller and thus the chances of actually 
producing an ultra-compact system decrease.
\end{itemize}

If we use a slightly more sophisticated, saturated magnetic braking law, the results are qualitatively 
similar to decreasing the magnetic braking strength. 
Because of the different dependencies of the two different magnetic braking laws on 
the mass and radius of the star in Eqs.\,\ref{eq:magnbrak} and \ref{eq:satMB}, the two prescriptions
take effect at completely different parts of the evolution.  To illustrate this, we picked
three models with an initial mass of $1\,M_\odot$ that evolve to the same minimum period 
(28\,min) at about the same mass (0.06--0.07 M$_\odot$).
The three models have different magnetic braking laws implemented and are given different initial periods to reach the 
desired $P_\mathrm{min}$: the first model uses braking according to Eq.\,\ref{eq:magnbrak} and 
$P_\mathrm{i} = 1.485$\,d so that the period minimum is reached after 11.7\,Gyr. The second model 
loses angular momentum according to the saturated magnetic braking law in Eq.\,\ref{eq:satMB}.  It 
has $P_\mathrm{i} = 1.109$\,d and needs 20.1\,Gyr.  The third model has no magnetic braking but only gravitational
waves to lose angular momentum. It needs the shortest initial period (0.4998\,d) and longest evolution
time (42\,Gyr) to reach the desired minimum period.

\begin{figure}
\resizebox{\hsize}{!}{\includegraphics{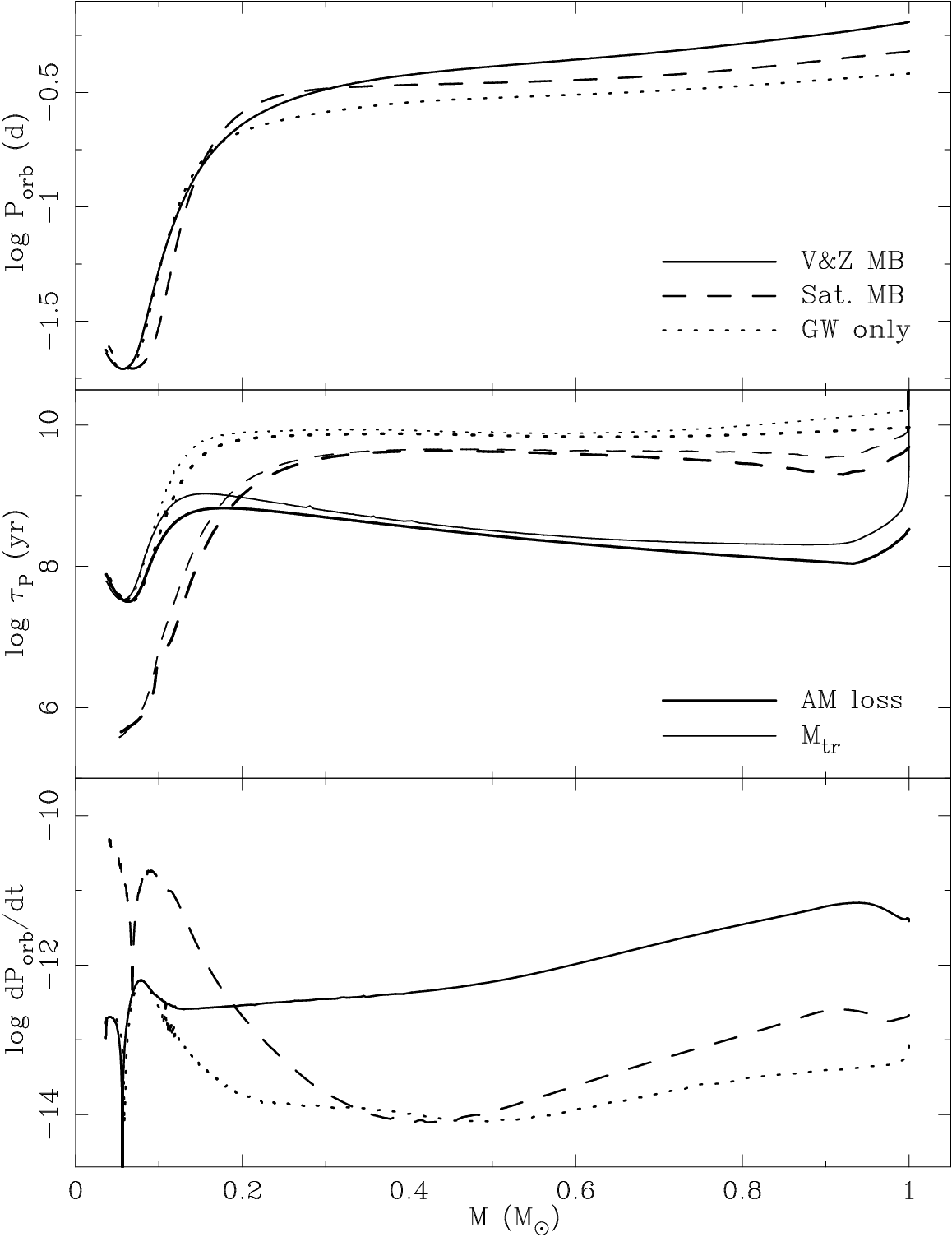}}
 \caption{ 
 {\it Upper panel} {\bf a)}: The logarithm of the orbital period.  The 
 line styles show the different models: with magnetic braking according to 
 Eq.\,\ref{eq:magnbrak} (V\&Z MB, solid line), with magnetic braking according to 
 Eq.\,\ref{eq:satMB} (Sat. MB, dashed line) and without magnetic braking (GW only, dotted line).
 {\it Middle panel} {\bf b)}: The logarithm of the timescales of orbital 
 shrinkage due to angular momentum loss (AM loss, thick lines) and orbital 
 expansion due to the mass transfer (M$_\mathrm{tr}$, thin lines).  The line styles represent
 the different models as in a).
 {\it Lower panel} {\bf c)}: The logarithm of (the absolute value of)
 the orbital period derivative in dimensionless units.  The line styles 
 are as in a).
 \label{fig:dpdt}  }
\end{figure}

The three models are compared in Fig.\,\ref{fig:dpdt}.  The data are shown as a 
function of the total mass of the donor, starting with the onset of Roche-lobe 
overflow.  Fig.\,\ref{fig:dpdt}a displays the orbital evolution of the three models.
Due to loss of angular momentum, the orbital periods at the start of mass transfer 
are already significantly shorter than the periods $P_i$ at the ZAMS.  
The model with the magnetic braking law of Eq.\,\ref{eq:magnbrak} has the
longest orbital period at the onset of RLOF, but shrinks fast and coincides with
the model without magnetic braking in the end.  The dashed line of the model with the saturated 
magnetic braking from Eq.\,\ref{eq:satMB} intersects the solid line twice before the period minimum,
indicating that braking starts out weaker, but ends stronger than the canonical magnetic braking of
Eq.\,\ref{eq:magnbrak}.
This is clearly seen in Fig.\,\ref{fig:dpdt}b, where for each model two competing
time scales are plotted: the time scale in which the mass transfer from the less
massive to the more massive component would expand the orbit if it were the only
process going on, and the timescale in which angular momentum loss (the sum of
gravitational radiation and magnetic braking) would shrink the orbit if nothing 
else would happen.  
In order to obtain the timescale at which the orbital period changes ($\tau_P$)
due to angular momentum loss ($\dot{J}$), we use the fact that the total 
angular momentum of a binary scales with the cubed root of the orbital period 
($J_\mathrm{orb} \propto P_\mathrm{orb}^{1/3}$) and thus
\begin{equation}
 \tau_P ~=~ \frac{P_\mathrm{orb}}{\dot{P}_\mathrm{orb}} ~=~ \frac{P_\mathrm{orb}}{\frac{dP}{dJ}\,\dot{J}_\mathrm{orb}} ~=~ \frac{J_\mathrm{orb}}{3\,\dot{J}_\mathrm{orb}}.
 \label{eq:taup}
\end{equation}
To calculate $\tau_P$ due to angular momentum loss we substitute for 
$\dot{J}_\mathrm{orb}$ in Eq.\,\ref{eq:taup} the sum of 
the angular momentum losses due to gravitational radiation and magnetic braking.
The period derivative due to conservative mass transfer from star 1 to star 2, assuming no angular momentum loss, is
given by:
\begin{equation}
  \dot{P}_\mathrm{orb} ~=~ 3P_\mathrm{orb}~\frac{M_1-M_2}{M_1\,M_2} ~\dot{M}_1 
\end{equation}
which can be substituted into Eq.\,\ref{eq:taup} to get $\tau_P$.
Depending on which of the two timescales is shorter, the
orbit will expand or shrink.  At the period minimum, the two lines for each model
intersect.  The figure shows that the timescales for the model with the 
canonical magnetic braking and the model with gravitational radiation only coincide around the period minimum. 
This happens because at these short orbital periods the models with canonical 
braking have very weak magnetic braking due to their small masses and radii 
(see Eq.\,\ref{eq:magnbrak}) and therefore gravitational wave emission dominates the orbital 
evolution.
It can be clearly seen in the figure that the timescales
for the model with saturated magnetic braking are almost two orders of magnitude shorter than
for the two other models and the orbital evolution is driven by the strong magnetic
braking.
Figure\,\ref{fig:dpdt}c shows the true period derivatives of the three models,
which could have been inferred from the difference between the lines in Fig.\,\ref{fig:dpdt}b.
It shows clearly that the orbit changes fastest for the model with canonical magnetic braking in the
first part of the evolution, but faster for the model with the saturated magnetic braking law when
the donor mass drops below about 0.2\,$M_\odot$.  Interestingly, the model with saturated magnetic
braking is in the saturated regime during all of the evolution, so that the difference in strength comes from the
different dependence on the mass and radius of the donor.  The deep dips in Fig.\,\ref{fig:dpdt}c 
are the period minima where $\dot{P}$ changes sign.
Figure\,\ref{fig:dpdt} illustrates that the two magnetic braking prescriptions
that we use work at completely different phases of the evolution of the model.  The
canonical braking model of Eq.\,\ref{eq:magnbrak} acts mainly in the first part of the
mass transfer phase, well before the period minimum, up to the point where the orbital period
has decreased enough for gravitational radiation to take over as the main angular momentum
loss mechanism and evolve the orbit to the ultra-short period regime.  The saturated magnetic
braking prescription of Eq.\,\ref{eq:satMB} is only slightly stronger than the gravitational
radiation in the first part of the evolution, but becomes orders of magnitude stronger at 
shorter orbital periods and evolves to the ultra-compact binary state without any significant
contribution in the angular momentum loss from gravitational radiation.  Despite these large
differences, there is little influence on the outcome of our statistical study.  We therefore
conclude that our study is independent of the the details of the magnetic braking, and that
the use of other theoretical or semi-empirical laws mentioned in Sect.\,\ref{sec:code} will
lead to similar results.


\section{Conclusions}
\label{sec:concl}

In our previous article we showed that for magnetic braking according to 
\citet{1981A&A...100L...7V} the formation of ultra-short-period binaries via
magnetic capture is possible, albeit very improbable, within the Hubble time.
In this paper we find that for less strong magnetic braking, in better agreement
with recent observations of single stars, the formation of ultra-short-period binaries via
magnetic capture is even less efficient.  Specifically, for magnetic braking reduced to 25\%
of the standard prescription (according to Eq.\,\ref{eq:magnbrak}), the shortest possible period
is 23\,min; for saturated magnetic braking (according to Eq.\,\ref{eq:satMB}) the shortest 
possible period is essentially the same as without magnetic braking, about 70\,min.

Loss of mass and associated angular momentum from the binary in general widens the orbit 
and thereby delays the formation of ultra-compact binaries.  However, this effect is
only noticeable in the absence of magnetic braking.

An attractive feature of the magnetic capture model is its ability to explain the negative
period derivative of the 11-minute binary in the globular cluster 
NGC\,6624  \citep{1993A&A...279L..21V, 2001ApJ...563..934C}.
Since we find that for a more realistic magnetic braking law it is impossible
to create ultra-compact binaries via magnetic capture at all, it becomes less 
likely that the negative period derivative is intrinsic.
\citet{1993MNRAS.260..686V} show that an apparent negative period derivative
can be the result of acceleration of the binary in the cluster potential.
According to measurements with the HST 
the projected position of the binary is very close to the cluster centre,
which makes a significant contribution of gravitational acceleration to the observed
period derivative more likely \citep{1993ApJ...413L.117K}.



\bibliographystyle{aa}
\bibliography{2696}

\end{document}